\documentclass[aps,prl,twocolumn,showpacs,groupedaddress,superscriptaddress]{revtex4}
\usepackage{graphicx,graphics,times}
\begin{document}




\textbf{Comment on "Quantum Key Distribution in the Holevo
Limit"}\\

 In a Letter, Cabello \cite{Cab} proposed a quantum key
distribution (QKD) Protocol which attended to Holevo limit. He
assumed that the quantum channel is composed of two qubits (1 and
2), and is prepared with equal probabilities in one of four
orthogonal pure states ${|\psi_{i}\rangle}$ ($i=0,1,2,3$), and
that Eve cannot have access to qubit 2 while she still holds the
qubit 1. He considers the following states (for simplicity, we
have replaced polarization states with spin states)
\begin{eqnarray}
&|\psi_{0}\rangle=|00\rangle_{12}\hspace{.5cm}
|\psi_{1}\rangle=\frac{1}{\sqrt{2}}[|10\rangle_{12}+|01\rangle_{12}]\nonumber\\
&|\psi_{2}\rangle=\frac{1}{\sqrt{2}}[|10\rangle_{12}-|01\rangle_{12}]\hspace{.5cm}
|\psi_{3}\rangle=|11\rangle_{12}
\end{eqnarray}
We use the states $|0\rangle$ and $|1\rangle$ to represent spin up
and spin down respectively. The \text{efficiency} of a QKD
 protocol, $E$, is defined as \cite{Cab}:
\begin{eqnarray}
E=\frac{b_{s}}{q_{t}+b_{t}},
\end{eqnarray}
where $b_{s}$ is the expected number of secret bits received by
Bob, $q_{t}$ is the number of qubits used in the quantum channel,
and $b_{t}$ is the number of bits used in the classical channel
between Alice and Bob. In Cabllo's scheme $b_{t}=0$ and
$q_{s}=q_{t}$, which leads to $E=1$ (the Holevo limit). Cabello's
protocol (CP) has some basic properties: (a) CP uses all of
Hilbert space dimensions, (b) He has defined an interesting
criterion for the efficiency of QKD protocols, (c) He avoids using
classical channel.

Eve could use a simple plan to distinguish between $(\psi_{0},
\psi_{3})$ and $(\psi_{1}, \psi_{2})$, without being detected by
Alice and Bob. In other words, the set of four states are now
partitioned into two sets. To show this, we assume Eve's particle
to be in the state $|0\rangle_{e}$. When the qubit 1 passes
through the first channel, she applies a \textsc{cnot} operation
on the qubit 1, as the control qubit, and her own particle, as the
target. Then, she lets the qubit 1 goes to Bob's system. When she
receives the qubit 2, she does the same operation on it and on her
own qubit. After this operations, we have:
\begin{eqnarray}
&|\psi_{i}\rangle|0\rangle_{e}\longrightarrow|\psi_{i}\rangle|0\rangle_{e}
\hspace{1cm} i=0,3
\nonumber\\&|\psi_{i}\rangle|0\rangle_{e}\longrightarrow|\psi_{i}\rangle|1\rangle_{e}
\hspace{1cm} i=1,2
\end{eqnarray}
At this stage, there is a trick by which Eve can distinguish
exactly between two of the four states \cite{Lon}. After the
application of the second \textsc{cnot}, she makes a measurement
on her own particle in the $\{{|0\rangle, |1\rangle}\}$ basis. If
she gets $|1\rangle$, she would let the qubit 2 go to Bob's
system. Otherwise, by a measurement on the qubit 2 in the
$\{{|0\rangle, |1\rangle}\}$ basis, she could understand whether
the state of Alice and Bob is $|00\rangle$ or $|11\rangle$. This
means that an undetectable Eve can know Alice's encoding whenever
Alice uses the basis $(\psi_{0}, \psi_{3})$, making the CP
insecure.

In the BB84 \cite{BB84} protocol, security is guaranteed by the
no-cloning theorem in the non-orthogonal states. CP was founded on
the no-cloning principle for orthogonal states, which was
suggested by T. Mor \cite{Mor}. He proposed that the two (or more)
orthogonal states cannot be cloned, if the reduced density
matrices of the first subsystem are non-orthogonal and
non-identical and the reduced density matrices of the second
subsystem are non-orthogonal. Here a basic question is whether
no-cloning condition for non-orthogonal and orthogonal states is
sufficient to have the security of QKD protocol? With attention to
our Eavesdropping approach, it seems that Mor's arguments for
no-cloning principal for orthogonal states \cite{Mor} is not
general enough to avoid eavesdropping. In what follows, we would
like to show that our approach is not restricted to CP. For
example, we consider two non-maximally entangled states as
follows:
\begin{eqnarray*}
|\psi\rangle=\cos\alpha|0\rangle_{1}|1\rangle_{2}+\sin\alpha|1\rangle_{1}|0\rangle_{2}
\\
|\phi\rangle=\cos\beta|0\rangle_{1}|0\rangle_{2}+\sin\beta|1\rangle_{1}|1\rangle_{2}
\end{eqnarray*}
with $0 <\alpha, \beta <\pi/2$ and $\alpha\neq\beta\neq \pi/4$.
The reduce density matrices for the first subsystems are:
$\rho^{1}_{\psi}=\cos^{2}\alpha|0\rangle\langle0|+\sin^{2}\alpha|1\rangle\langle1|,
\varrho^{1}_{\phi}=\cos^{2}\beta|0\rangle\langle0|+\sin^{2}\beta|1\rangle\langle1|$.
These reduced density matrices are neither orthogonal nor
identical, and so they could't be cloned, and the reduced density
matrices for second subsystems are:
$\rho^{2}_{\psi}=\cos^{2}\alpha|1\rangle\langle1|+\sin^{2}\alpha|0\rangle\langle0|,
\varrho^{2}_{\phi}=\cos^{2}\beta|0\rangle\langle0|+\sin^{2}\beta|1\rangle\langle1|$.
These reduced density matrices are not orthogonal to each other,
but this protocol is not secured completely against a double
\textsc{cnot} operations.

Here are some questions: What are the defects in the Mor proposal?
How Mor proposal can be repaired? In another paper in progress, we
respond to these questions.








The authors are grateful to referee for useful comments and A. T.
Rezakhani for critical reading.(this work is supported under
project name: \emph{Aseman}).
\\
A. Fahmi$^{1}$ and M. Golshani$^{1,2}$\\
fahmi@theory.ipm.ac.ir\\
$^{1}$School of Physics, Institute for Studies in Theoretical
Physics and Mathematics (IPM)
P. O. Box 19395-5531, Tehran, Iran\\
$^{2}$Department of Physics, Sharif University of Technology, P.
O. Box
11365-9161, Tehran, Iran.


\end{document}